\documentclass[aps,eqsecnum,preprint,preprintnumbers,12pt,amsfonts]{revtex4}

\usepackage{epsfig,psfrag}
\usepackage{latexsym}

\def\beq{\begin{eqnarray}}
\def\eeq{\end{eqnarray}}
\def\bea{\begin{eqnarray}}
\def\eea{\end{eqnarray}}
\def\det{{\rm det}}
\newcommand{\nn}{\nonumber}
\newcommand{\reals}{\mbox{${\rm I\!R }$}}
\newcommand{\nats}{\mbox{${\rm I\!N }$}}
\newcommand{\intgs}{\mbox{${\rm Z\!\!Z }$}}

\newcommand{\intl}{\int\limits}

\newcommand{\intv}{\int\limits_{-\infty}^\infty dx \,\,}
\newcommand{\kpm}{k^2-p_n^2-m^2}
\newcommand{\pmq}{p_n^2+m^2}
\newcommand{\snmp}{\sum_{n=-\infty}^\infty}
\newcommand{\sln}{\sum_{l=0}^\infty}
\newcommand{\pmr} {\left( 1+\frac{(\pmq )r^2}{\nu^2}\right)}
\newcommand{\pmrr}{\left( 1 + \frac{ m^2 r^2}{\nu^2 + p_n^2
r^2}\right)}
\newcommand{\pr}{\left( 1 + \frac{p_n^2 r^2} {\nu^2}\right)}

\newcommand{\we}{\left(\frac{2\pi}{L_1}\right)^2}
\newcommand{\wz}{\left(\frac{2\pi}{L_2}\right)^2}
\newcommand{\wdd}{\left(\frac{2\pi}{L_3}\right)^2}

\usepackage{color}

\begin{document}

\title{Simplified Vacuum Energy Expressions for  Radial Backgrounds and Domain Walls}
\author{Gerald V. Dunne}\email{dunne@phys.uconn.edu}
\affiliation{Department of Physics, University of Connecticut,
Storrs, CT 06269}
\author{Klaus Kirsten}\email{klaus_kirsten@baylor.edu}
\affiliation{Department of Mathematics, Baylor University, Waco,
TX 76798}


\begin{abstract}
We extend our previous results of simplified expressions for functional determinants for radial
Schr\"odinger operators to the computation of vacuum energy, or mass corrections, for static but spatially
radial backgrounds, and for domain wall configurations. Our method is based on the zeta function approach to the Gel'fand-Yaglom theorem, suitably extended to higher dimensional systems on separable manifolds. We find new expressions that are easy to implement numerically, for both zero and nonzero temperature.
\end{abstract}

\maketitle

\section{Introduction}
\label{sec-intro}

Determinants of differential operators occur naturally in many
applications in mathematical and theoretical physics, and also
have inherent mathematical interest since they encode certain
spectral properties of differential operators. Physically, such
determinants arise in semiclassical and one-loop approximations in
quantum mechanics and quantum field theory
\cite{sala53-90-690,schw54-94-1362,sala74-9-1129,jack74-9-1686,ilio75-47-165,cole79u,dunn08-41-304006}.
Determinants of free Laplacians and free Dirac operators have been
extensively studied
\cite{ray71-7-145,hawk77-55-133,eliz94b,byts96-266-1,dhok86-104-537,dowk76-13-3224,sarn87-110-113,kirs02b},
but much less is known about operators involving an arbitrary
potential function. For ordinary ({\it i.e.}, one dimensional)
differential operators, a general theory has been developed for
determinants of such operators
\cite{gelf60-1-48,levi77-65-299,form87-88-447,kirs02b,kirs03-308-502,kirs04-37-4649,klei06b}.
In this paper we extend these results to a broad class of
separable {\it partial} differential operators.
Our  approach is based on the zeta function definition of the determinant \cite{ray71-7-145,hawk77-55-133}, together with dimensional regularization and renormalization. Thus, it is similar in spirit to some previous approaches \cite{wipf,baacke1993,bord00-61-085008,bord96-53-5753,isidori2001,shifman1998,parnachev2000,graham2001,graham2002,graham-olum,rebhan2002,baacke2004,dunne2007}, but our main new result is that we are able to simplify the final expressions considerably by systematically separating out finite parts of the subtractions.

Let $x\in\reals$ and $y\in M$, with $M$ some $(d-1)$ dimensional
manifold. We consider functional determinants of Laplace type
operators of the form
\beq
L = - \frac {d^2}{dx^2} - \Delta_y +
m^2 + V (x) + W (y) \quad .
\label{1}
\eeq
Using a separation of
variables, $$ \phi (x,y) = X (x) Y (y),$$ the eigenvalues and
eigenfunctions for this operator are determined by \beq \left( -
\frac {d^2}{dx^2} + m^2 + \lambda^2 + V(x) \right) X (x) = k^2 X
(x) , \nn\eeq where \beq\left( -\Delta_y  + W (y)  \right) Y(y) =
\lambda^2  Y(y) .\nn\eeq

In this paper we consider two situations in which we can obtain computationally simple expressions for the determinant of $L$.

(i) The first case is where $V(x)=0$, and $W(y)$ defines a radial
potential on the $(d-1)$ dimensional space $M$. The physical
application motivating this case is the computation of induced
vacuum energy, or one-loop mass corrections, in quantum field
theory for backgrounds that are {\it static} [we regard $x$ as the
Euclidean time coordinate] and spatially radially symmetric [$M$ is the
spatial manifold]. For a bosonic field the vacuum energy is given
by \beq \Delta E= -\frac{1}{2}{\rm tr}_p \log\,\det\left(
-\Delta_y  + W (y)+m^2+p^2\right) \label{energy} \eeq where ${\rm
tr}_p$ is a sum over Matsubara modes at finite temperature, or an
integral at zero temperature \cite{kapu89b}: \beq {\rm tr}_p\,
g(p)\equiv \cases{ \frac{1}{\beta} \sum_{n=-\infty}^\infty  g(p_n)
\quad, \quad p_n= \frac{ 2 \pi n}{\beta}  \quad, \quad {\rm finite
\, \,temperature, } \cr \int_{-\infty}^\infty \frac{dp}{2\pi} g(p)
\quad\quad , \quad {\rm zero\,\, temperature.}} \label{matsubara}
\eeq

(ii) The second case is where $V(x)$ is non-zero, and $M$ is a
non-trivial $(d-1)$-dimensional manifold such as a torus
$T^{d-1}$, or a sphere $S^{d-1}$, with $W(y)=0$. In this case, we study the effect of the nontrivial topology of $M$ on the determinant.

In each of these cases, there is a natural separation of variables, but the sum over the remaining
eigenvalues (partial waves, Matsubara modes, or Kaluza-Klein modes) is divergent and must be regularized and
renormalized. We solve these regularization and renormalization issues using the zeta function approach.

\section{Vacuum energy in a spherically symmetric background
field}
\label{sec-energy}

In this section we find a computationally simple expression for the vacuum energy
(\ref{energy}) for a bosonic field in a static but spatially radial background.
We work at finite temperature, but the limit of the final result to  zero temperature will be clear.
A possible first guess would be to take the result of
\cite{dunn06-39-11915} for a radial determinant in dimension
$(d-1)$, replace $m^2$ by $m^2+p_n^2$, and trace over the
eigenvalues $p_n$. However, while the finite renormalized [in
$(d-1)$ dimensions] determinant for any given $p_n$ can be
computed using the results of \cite{dunn06-39-11915}, the
subsequent trace over $p_n$ is divergent. This divergence contains
important physics and must be treated appropriately, namely renormalizing in $d$ dimensions is needed.
Our main result is equation (\ref{4d-answer}) below, which expresses the finite and renormalized log determinant in $(3+1)$-dimensions in a form that is easy to compute numerically.

We begin with the zeta function, and by standard techniques
\cite{bord96-53-5753,kirs02b}, we express the zeta function in
terms of the Jost function $f_l$ as:
 \beq
 \zeta (s) = \frac{\sin \pi s}{\pi\beta} \snmp \sln {\rm deg}(l; d-1)
\intl_{\sqrt {\pmq }}^\infty dk \,\, \left( \kpm \right)^{-s}
\frac d {dk} \ln f_l (ik) \quad , \label{zeta} \eeq where ${\rm
deg}(l; d-1)$ is the degeneracy of the $l^{\rm th}$ partial wave
for the $(d-1)$ dimensional radial Laplacian: \beq {\rm deg}(l;
d-1)= \frac{(2l+d-3)(l+d-4)!}{l!(d-3)!} .\label{deg} \eeq The task
is to find the analytic continuation of this zeta function so that
it and its derivative can be analyzed in the neighborhood of
$s=0$. Such an analytic continuation can be achieved by adding and
subtracting the uniform asymptotic behavior, $f_l^{asym}(ik)$, of
the Jost function $f_l (ik)$. This asymptotic behavior is well
known from scattering theory \cite{tayl72b}: \beq \ln
f_l^{asym}(ik) &\sim& \frac 1 {2\nu} \intl_0^\infty dr \,\,
\frac{r\,
W(r)} {\left( 1 + \frac {k^2 r^2} {\nu^2} \right)^{1/2}} \nn  \\
& &\hspace{-3.0cm}+ \frac 1 {16\nu^3} \intl_0^\infty dr \,\,
\frac{r\,W(r)} {\left( 1 + \frac {k^2 r^2} {\nu^2} \right)^{3/2}} \left[ 1 - \frac{6}{\left( 1 + \frac {k^2 r^2} {\nu^2} \right)}
+ \frac{5}{\left( 1 + \frac {k^2 r^2} {\nu^2} \right)^2} \right]
\nn\\
& &\hspace{-3.0cm}-\frac 1 {8\nu^3} \intl_0^\infty dr \,\,
\frac{r^3\, W^2 (r)}{\left( 1 + \frac {k^2 r^2} {\nu^2}
\right)^{3/2}} +\dots \label{jost-asymptotic} \eeq where for a
$(d-1)$ dimensional radial system \beq \nu\equiv l+\frac{d-1}{2}-1
\quad . \label{nu} \eeq Since we are interested in renormalizable
theories with $d\leq 4$, it is sufficient to consider this many
terms in the expansion \footnote{A natural mathematical
generalization of a finite determinant in any dimension is
possible \cite{simo77-24-244}, but requires higher order terms in
this asymptotic expansion.}. Eventually, the convergence of our final  expressions
could be accelerated even more using
further subtractions, as discussed in \cite{hur2008} for effective actions, but we do not pursue this here.
For $d=3$ we must first separate out
the $l=0$ term. We first concentrate on $d=4$, and so $\nu\neq 0$.

Therefore, {\it defining} the function $\ln f_l^{asym}(ik)$ to be
just the terms shown in (\ref{jost-asymptotic}), we are naturally
led to the splitting: \beq \zeta (s) = \zeta _f (s) + \zeta _{as}
(s) \label{split} \eeq
 with the definitions
 \beq \zeta_f (s) &=&
\frac{\sin \pi s} {\pi\beta} \snmp \sln {\rm deg}(l; d-1) \intl_{\sqrt{\pmq}}^\infty
dk \,\, (\kpm )^{-s} \frac d {dk} \left[ \ln f_l (ik) - \ln
f_l^{asym} (ik)\right] ,\nn\\
 \label{zeta-f}\\
\zeta _{as} (s) &=&\frac{\sin \pi s} {\pi\beta} \snmp \sln {\rm
deg}(l; d-1) \intl_{\sqrt{\pmq}}^\infty dk \,\, (\kpm )^{-s} \frac
d {dk} \ln f_l^{asym} (ik) .\label{zeta-as} \eeq Then, by
construction, $\zeta_f (s)$ is well defined about $s=0$, and
$\zeta^\prime_f(0)$ is finite and given by
 \beq
 \zeta _f ' (0) = - \frac{1}{\beta}\snmp \sln  {\rm deg}(l; d-1) \left[ \ln f_l ( i
\sqrt{ \pmq}) - \ln f_l ^{asym} (i \sqrt{ \pmq})\right] \quad .
\label{zeta-prime-f}
\eeq
Likewise, $\zeta_{as}(s)$ can be analytically continued to the neighborhood of $s=0$, and its
derivative  $\zeta_{as}^\prime(0)$ can be defined there.

Our main observation is that the final formula that results from
this procedure, as reported in
\cite{bord96-53-5753,bord00-61-085008}, may be dramatically
simplified owing to cancellations of finite terms between
$\zeta_{f}^\prime(0)$ and $\zeta_{as}^\prime(0)$. These finite
terms can be cancelled at the beginning, leading to much simpler
expressions for each of $\zeta_{f}^\prime(0)$ and
$\zeta_{as}^\prime(0)$, with their sum being unchanged.
This leads to an expression that is precisely equal, but is significantly easier to compute numerically.

\subsection{Rewriting $\zeta^\prime_f(0)$}

Our procedure for rewriting $\zeta^\prime_f(0)$  is based on the simple observation that
we can write
\beq
\pmr = \pr \pmrr
\label{identity}
\eeq
and the second factor can now be expanded at both large partial wave number (i.e., large $l$, or large $\nu$),
and large $p_n$.
Performing a large-$l$
expansion directly on the left-hand-side is not possible
because $( \pmq )r^2/\nu^2$ is not a small parameter, as $p_n$ can
become arbitrarily large.

Proceeding in this fashion we can separate  $\ln\, f_l^{asym} (i\sqrt{p_n^2+m^2})$ according to
the order of its terms at large $(\nu^2+p_n^2 r^2)$:
\beq
\ln\, f_l^{asym} (i\sqrt{p_n^2+m^2}) \equiv \ln\, f_l^{asym, (1)} (i\sqrt{p_n^2+m^2})+\ln\, f_l^{asym, (2)} (i\sqrt{p_n^2+m^2})
\label{zeta-f-split}
\eeq
where
\beq
\ln\, f_l^{asym, (1)} (i\sqrt{p_n^2+m^2}) &=& \frac 1 {2\nu}
\intl_0^\infty \frac{r W(r)} {\pr ^{1/2}} - \frac 1 {8 \nu^3}
\intl_0^\infty dr \,\, \frac{r^3 W(r) (W(r) + 2m^2) }{\pr ^{3/2}}
\label{zeta-f-1}\\
& &+\frac 1 {16 \nu^3} \intl_0^\infty dr \,\, \frac{r W(r)}
{\pr ^{3/2}} \left[ 1-\frac 6 { \pr} + \frac 5 {\pr ^2} \right]
\nn
\eeq
contains terms of ${\cal O} ((\nu^2 + p_n^2 r^2)^{-1/2})$ and ${\cal O} ((\nu^2+p_n^2 r^2)^{-3/2})$, while
\beq
\ln\, f_l^{asym, (2)} (i\sqrt{p_n^2+m^2}) &=&
\frac 1 {2\nu} \intl_0^\infty dr \,\, \frac {r W(r)} {\pr
^{1/2}} \left[ \frac 1 {\pmrr ^{1/2}} - 1 + \frac 1 2 \frac{m^2
r^2}{\nu^2 + p_n^2 r^2} \right] \nn\\
& &- \frac 1 {8 \nu^3} \intl_0^\infty dr \,\, \frac{r^3 W^2(r)}
{\pr ^{3/2}} \left[ \frac 1 {\pmrr ^{3/2}} - 1 \right] \nn\\
& &+\frac 1 {16 \nu^3} \intl_0^\infty dr \,\, \frac{r W(r)}
{\pr ^{3/2}} \left[ \frac 1 {\pmrr ^{3/2}} - 1 \right] \nn\\
& &-\frac 3 {8 \nu^3} \intl_0^\infty dr \,\, \frac{r W(r)}
{\pr ^{5/2}} \left[ \frac 1 {\pmrr ^{5/2}} - 1 \right] \nn\\
& &+\frac 5 {16 \nu^3} \intl_0^\infty dr \,\, \frac{r W(r)} {\pr
^{7/2}} \left[ \frac 1 {\pmrr ^{7/2}} - 1 \right]
\label{zeta-f-2}
\eeq
contains terms of order ${\cal O} ((\nu^2 + p_n^2 r^2)^{-5/2})$. Thus, for $\ln\, f_l^{asym, (2)} (i\sqrt{p_n^2+m^2})$, the
 $l$- and $n$-summations in (\ref{zeta-prime-f}) are finite. It is therefore
sufficient to subtract only $\ln\, f_l^{asym, (1)} (i\sqrt{p_n^2+m^2})$  from $\ln f_l (i
\sqrt {\pmq} )$ to obtain a finite answer. Thus, we rewrite (\ref{zeta-prime-f}) as
 \beq
 \zeta _f ' (0) &=& - \frac{1}{\beta}\snmp \sln  {\rm deg}(l; d-1) \left[ \ln f_l ( i
\sqrt{ \pmq}) - \ln f_l ^{asym, (1)} (i \sqrt{ \pmq})\right] \nn\\
&&
 + \frac{1}{\beta}\snmp \sln  {\rm deg}(l; d-1) \left[ \ln f_l ^{asym, (2)} (i \sqrt{
 \pmq})\right].
\label{zeta-prime-f-new}
\eeq
By construction, the summations are {\it separately finite}, and the total is precisely equal to the expression
in (\ref{zeta-prime-f}). The main rationale behind this splitting is that we show below that the entire contribution of the sum over $\ln\, f_l^{asym, (2)} (i\sqrt{p_n^2+m^2})$ in (\ref{zeta-prime-f-new}),
which is completely finite, is cancelled precisely by identical terms coming from $\zeta_{as}^\prime(0)$. It is therefore computationally redundant to compute these terms.

\subsection{Rewriting $\zeta^\prime_{as}(0)$}

A similar strategy is applied to the evaluation of
$\zeta^\prime_{as}(0)$. We begin with the definition
(\ref{zeta-as}) of the zeta function $\zeta_{as}(s)$, and perform
the $k$ integrations using the basic dimensional regularization
formula \beq \intl_{\sqrt {\pmq}}^\infty dk \,\, (\kpm )^{-s}
\frac d {dk} \left( 1 + \frac{k^2 r^2} {\nu^2} \right) ^{-\frac N
2} = - \frac{\pi\,\Gamma \left( s+\frac N 2 \right) }{\sin(\pi
s)\,\Gamma \left( \frac N 2 \right)\Gamma (s)} \frac{r^{2s}
\nu^{-2s}}{\pmr ^{s+N/2}}.
\nn\\
\label{k-int} \eeq Referring to (\ref{jost-asymptotic}), we see
that each term in $\zeta_{as}(s)$ is of the form where this
integration formula can be applied. Therefore, the $k$ integrals in (\ref{zeta-as}) produce a set of terms
of the form \beq {\pmr ^{s+N/2}} \eeq for  $N=1, 3, 5, 7$. We now
make an expansion based on the decomposition in (\ref{identity}),
which naturally separates $\zeta_{as}(s)$ into two parts,
mirroring the separation of $\ln f_l^{asym}(i\sqrt{p_n^2+m^2})$ in
(\ref{zeta-f-split}): \beq \zeta_{as}(s)\equiv
\zeta_{as}^{(1)}(s)+\zeta_{as}^{(2)}(s) \quad ;
\label{zeta-as-split} \eeq for details see Appendix A. The first
term, $\zeta_{as}^{(1)}(s)$, is \beq
 \zeta_{as}^{(1)}(s)&=& \frac{1}{\beta}\sum_{n=-\infty}^\infty\sum_{l=0}^\infty{\rm deg}(l; d-1)\left\{ -\frac{1}{2}\frac{\Gamma\left(s+\frac{1}{2}\right)}{\Gamma\left(\frac{1}{2}\right)\Gamma\left(s\right)}\int_0^\infty dr
\frac{\frac{r^{2s+1}}{\nu^{2s+1}} W(r)} {\pr ^{s+1/2}} \right.\nn\\
&&\left. + \frac{1}{8}\frac{\Gamma\left(s+\frac{3}{2}\right)}{\Gamma\left(\frac{3}{2}\right)\Gamma\left(s\right)}
\intl_0^\infty dr \,\, \frac{\frac{r^{2s+3}}{\nu^{2s+3}} W(r) (W(r) + 2m^2) }{\pr ^{s+3/2}}\right.
\label{zeta-as-1}\\
& &\left. -\frac 1 {16
\nu^2}\frac{\Gamma\left(s+\frac{3}{2}\right)}{\Gamma\left(\frac{3}{2}\right)\Gamma\left(s\right)}
\intl_0^\infty dr \,\, \frac{\frac{r^{2s+1}}{\nu^{2s+1}} W(r)}
{\pr ^{s+3/2}} \left[ 1-\frac{4\left(s+\frac{3}{2}\right)}{ \pr} +
\frac{\frac{4}{3}
\left(s+\frac{3}{2}\right)\left(s+\frac{5}{2}\right)} {\pr ^2}
\right]\right\}. \nn \eeq The second term in the decomposition
(\ref{zeta-as-split}) is \beq \zeta_{as}^{(2)}(s) &=&
\frac{1}{\beta}\frac{\Gamma
\left(s+\frac{1}{2}\right)}{\sqrt{\pi}\Gamma\left(s\right)}
\sum_{n=-\infty}^\infty\sum_{l=0}^\infty {\rm deg}(l; d-1)\times\label{zeta-as-2}\\
& &\left\{ -\frac{1}{2} \intl_0^\infty dr \,\, \frac
{\frac{r^{2s+1}}{\nu^{2s+1}} W(r)} {\pr ^{s+1/2}} \left[ \frac 1
{\pmrr ^{s+1/2}} - 1 +\left(s+\frac 1 2\right) \frac{m^2
r^2}{\nu^2 + p_n^2 r^2} \right]  \right.
\nn\\
& &\left .+ \frac {\left(s+\frac{1}{2}\right) }{4} \intl_0^\infty dr \,\, \frac{\frac{r^{2s+3}}{\nu^{2s+3}}  W^2(r)}
{\pr ^{s+3/2}} \left[ \frac 1 {\pmrr ^{s+3/2}} - 1 \right]  \right.
\nn\\
& &\left.-\frac{\left(s+\frac{1}{2}\right) } {8 \nu^2} \intl_0^\infty dr \,\, \frac{\frac{r^{2s+1}}{\nu^{2s+1}}  W(r)}
{\pr ^{s+3/2}} \left[ \frac 1 {\pmrr ^{s+3/2}} - 1 \right] \right.
\nn\\
& &
\left. +\frac{\left(s+\frac{1}{2}\right)\left(s+\frac{3}{2}\right)} {2 \nu^2} \intl_0^\infty dr \,\, \frac{\frac{r^{2s+1}}{\nu^{2s+1}} W(r)}
{\pr ^{s+5/2}} \left[ \frac 1 {\pmrr ^{s+5/2}} - 1 \right] \right.
\nn\\
& &\left.
-\frac{\left(s+\frac{1}{2}\right)\left(s+\frac{3}{2}\right)\left(s+\frac{5}{2}\right)}{6
\nu^2} \intl_0^\infty dr \,\, \frac{\frac{r^{2s+1}}{\nu^{2s+1}}
W(r)} {\pr ^{s+7/2}} \left[ \frac 1 {\pmrr ^{s+7/2}} - 1 \right]
\right\}. \nn \eeq It is clear that, by construction,
$\zeta_{as}^{(2)}(s)$ is regular at $s=0$, and a simple
computation yields \beq
\left[\frac{d}{ds}\zeta_{as}^{(2)}(s)\right]_{s=0}=-\frac{1}{\beta}\snmp
\sln  {\rm deg}(l; d-1) \left[ \ln f_l ^{asym, (2)} (i \sqrt{
\pmq})\right] \label{zeta-prime-as-2} \eeq so that it cancels
exactly the second, finite, sum in (\ref{zeta-prime-f-new}) for
$\zeta^\prime_f(0)$. Thus, we only need to compute
$\left[\frac{d}{ds}\zeta_{as}^{(1)}(s)\right]_{s=0}$.

To compute $\left[\frac{d}{ds}\zeta_{as}^{(1)}(s)\right]_{s=0}$,
it is useful to define the Epstein-like zeta function \beq
E_d^{(k)} (s,a) &\equiv& \frac{1}{\beta}\snmp \sln {\rm deg}(l;
d-1) \frac{\nu^k}{(\nu^2 + a^2 n^2)^s}. \label{epstein} \eeq Then
\beq \zeta_{as}^{(1)}(s)&=&- \frac{\Gamma \left( s + \frac 1 2
\right)} {2\sqrt \pi\Gamma (s)} \intl_0^\infty dr \, W\,
r^{1+2s}\, E_d^{(0)} \left( s + \frac 1 2 ,
\frac{2\pi r} \beta\right) \nn\\
&&+ \frac{\Gamma\left(s+\frac{3}{2}\right)}{4\sqrt{\pi}\Gamma\left(s\right)}
\intl_0^\infty dr \, W \left(W+ 2m^2\right)\, r^{2s+3}\,  E_d^{(0)} \left( s + \frac 3 2 , \frac{2\pi r}
\beta \right)\nn\\
& &-
\frac{\Gamma \left( s + \frac 3 2 \right)} {8 \sqrt \pi\Gamma (s)}
\intl_0^\infty dr \,W\, r^{1+2s} \left[ E_d^{(0)} \left( s + \frac 3 2 , \frac{2\pi r}
\beta\right) -4\left(s+\frac{3}{2}\right) E_d^{(2)} \left( s + \frac 52
, \frac{2\pi r}
\beta\right)  \right. \nn\\
& &\left. \hskip 2cm +
\frac{4}{3}\left(s+\frac{3}{2}\right)\left(s+\frac{5}{2}\right)E_d^{(4)}
\left( s + \frac 7 2 , \frac{2\pi r} \beta\right)\right].
\label{zeta-as-22} \eeq To compute
$\left[\frac{d}{ds}\zeta_{as}^{(1)}(s)\right]_{s=0}$, we need the
behavior of the zeta functions $E_d^{(k)}(s, z)$ in the vicinity
of $s=1/2, 3/2, 5/2, 7/2$. This behavior is particularly simple in
the zero temperature limit, in which we can replace the sum over
Matsubara modes by an integral as in (\ref{matsubara})
\cite{kapu89b}.
Then the zeta functions in (\ref{zeta-as-22}) reduce to Hurwitz
zeta functions, see (\ref{relzeta}), (\ref{epsind4}) and
(\ref{relzed3}). For example, in $d=4$, where ${\rm deg}(l;
d-1)=2l+1\equiv2\nu$, we find at $T=0$ \beq
\zeta_{as}^{(1)}(s)&=&-\frac{1}{2\pi}\int_0^\infty
dr\,W\,r^{2s}\left(2^{2s-1}-1\right)\zeta_R(2s-1)
\label{zeta-as-zero}\\
&&+\frac{1}{4\pi}\int_0^\infty dr\,W(W+2m^2)\,r^{2s+2}\,s\,\left(2^{2s+1}-1\right)\zeta_R(2s+1)\nn\\
&&-\frac{1}{8\pi}\int_0^\infty
dr\,W\,r^{2s}\left[s-4s(s+1)+\frac{4}{3}s(s+1)(s+2)\right]\,\left(2^{2s+1}-1\right)\zeta_R(2s+1).
\nn \eeq We thus find the following simple expression for the
derivative at $s=0$: \beq
\left[\frac{d}{ds}\zeta_{as}^{(1)}(s)\right]_{s=0}&=&\frac{1}{24\pi}\left[\gamma+3\ln
2+12\zeta_R^\prime(-1)\right]\int_0^\infty dr\,W(r)
\label{zeta-as-1-final}\\
&&+\frac{1}{4\pi}\int_0^\infty
dr\,W(r)(W(r)+2m^2)\,r^{2}\,\left[\gamma+\ln(4\,r)\right]. \nn \eeq
Notice that this expression is very simple -- it does not involve
any summations.

\subsection{Combining the results for $\zeta^\prime_{f}(0)$ and $\zeta^\prime_{as}(0)$}

In the previous two subsections we showed that each of
$\zeta^\prime_{f}(0)$ and $\zeta^\prime_{as}(0)$ could be split
naturally into two parts, with all parts being manifestly finite,
and in such a way that the second part of $\zeta^\prime_{f}(0)$ in
(\ref{zeta-prime-f-new}) cancels against the second part of
$\zeta^\prime_{as}(0)$ in (\ref{zeta-prime-as-2}). Thus, we are
left with a much simpler expression for the net $\zeta^\prime(0)$:
\beq \left[\zeta^\prime(0)\right]_{d=4}&=& -\frac{1}{2 \pi
}\int\limits_{-\infty}^\infty dp \,\, \sum_{l=0}^\infty (2\nu)
\left[ \ln f_l (i\sqrt {p^2 +m^2}) - \frac 1 2\int\limits_0^\infty
dr \,\,\frac{r W}{(\nu^2+p^2 r^2)^{1/2}} \right. \label{4d-answer}
\\
& &\left.\hspace{-2.0cm}+\frac 1 8 \int\limits_0^\infty dr \,\,
\frac{r^3 W (W + 2m^2)}{(\nu^2+p^2r^2)^{3/2}} -\frac 1 {16}
\int\limits_0^\infty \frac{ dr\,r\, W(r)}{(\nu^2+p^2r^2)^{3/2}}
\left( 1-\frac{6\nu^2}{(\nu^2+p^2r^2)}+ \frac{5
\nu^4}{(\nu^2+p^2r^2)^2} \right) \right]\nn\\
& & \hspace{-2.0cm}+\frac{1}{24\pi}\left[\gamma+3\ln
2+12\zeta_R^\prime(-1)\right]\int_0^\infty dr\,W
+\frac{1}{4\pi}\int_0^\infty
dr\,W(W+2m^2)\,r^{2}\,\left[\gamma+\ln(4rm)\right] \nn \eeq where
$\nu=l+1/2$ for $d=4$; the vacuum energy now is $\Delta E = (1/2)
\zeta ' (0)$.

The result for finite temperature follows by the replacement
$$\frac 1 {2\pi} \int\limits_{-\infty}^\infty dp \to \frac 1 \beta \sum_{-\infty}^\infty \quad\quad \mbox{and}
\quad \quad p\to p_n.$$ Furthermore, there are exponentially
damped contributions as $\beta \to \infty$, involving series over
Bessel functions as described in Appendix A.

Equation (\ref{4d-answer}) and its finite temperature version are
our main results in this section. It is worth stressing the
computational simplicity of this result in comparison to previous
expressions. For each Matsubara mode $p_n$, and each partial wave
$l$, the logarithm of the Jost function, $\ln f_l (i\sqrt {p_n^2
+m^2})$, can be evaluated simply and efficiently using the radial
version of the Gel'fand-Yaglom theorem, as described in
\cite{dunn08-41-304006}. Our result states that, with the
subtractions in (\ref{4d-answer}), the double sum over $n$ and $l$
converges. These subtraction terms are simple integrals, easy to
evaluate numerically. The finite counterterm on the last line does
not involve any summation, and can be evaluated once and for all
given the radial potential $W(r)$. By contrast, the corresponding
subtraction terms, and also the counterterms are significantly
more involved in \cite{bord96-53-5753}. The specific form of the
finite term on the last line of (\ref{4d-answer}) arises through
our use of dimensional regularization, and we have renormalized
on-shell at renormalization scale $\mu=m$. It is well-known how to
convert from a given renormalization scheme to another by finite
counter-terms \cite{bjor65b}, so our expression (\ref{4d-answer})
gives not just the finite determinant, but the finite and
renormalized vacuum energy.

In $d=3$ the $l=0$ mode needs separate treatment, the $l\geq 1$
modes are dealt with as before; see Appendix A for details. Again,
at $T=0$ the answer is surprisingly simple and reads \beq \zeta '
(0) &=&- \frac 1 {2\pi} \intl_{-\infty} ^\infty dp \left[ \ln f_0
(i\sqrt{p^2+m^2}) - \frac 1 {2\sqrt{p^2+m^2}}
\intl_0^\infty dr \,\, W(r) \right] \nn\\
& &-\frac 1 \pi \sum_{l=1}^\infty \intl_{-\infty}^\infty dp \left[
\ln f_l (i\sqrt{p^2+m^2}) - \frac 1 2 \intl_0^\infty dr\,\,
\frac{rW(r)} {(l^2 + p^2r^2)^{\frac 1 2}}\right]\nn\\
& &+\frac 1 {2\pi} \intl_0^\infty dr \,\, W(r) \ln (2\pi
rm).\nn\eeq The case $d=2$ follows from (\ref{6}) once we put $L=\beta$. For finite temperature the remarks below
(\ref{4d-answer}) remain valid.

\section{Compactification with nontrivial $V(x)$}
\label{vx}

We now turn to the opposite limit, in which the nontrivial
potential is not a radial potential on the $(d-1)$ dimensional
manifold $M$, but a one-dimensional potential for $x\in \reals$. Physically, this corresponds to
a domain wall configuration in the $x$ direction, with $M$ describing the transverse directions.
So let us assume $V(x)$ is the nontrivial potential with suitable
asymptotic properties as defined below in terms of the Jost
function. For the zeta function analysis we follow the one
dimensional scattering approach as described in
\cite{bord95-28-755}. In this calculation we assume $\lambda^2 >
0$; the contribution for $\lambda^2 = 0$ is easily added at the
end. Starting off as usual \cite{kirs02b} with a suitable contour
in the complex plane, the zeta function for the above operator,
after shifting it to the imaginary axis, reads \beq \zeta (s) = -
\frac{\sin (\pi s)} \pi \sum_\lambda \int\limits_{\sqrt{m^2 +
\lambda^2}}^\infty dp \left( p^2 - m^2 - \lambda ^2 \right) ^{-s}
\frac
\partial {\partial p } \ln s_{11} (ip) \label{2} \eeq with
$s_{11}$ a suitable element of the S-matrix. The asymptotics of
$s_{11} (ip)$ for large $p$ is known \cite{zakh72-5-280} and reads
\beq \ln s_{11} (ip) = - \frac 1 {2p} \int\limits_{-\infty}^\infty
dx \,\, V(x) + \frac 1 {8 p^3} \int\limits_{-\infty} ^\infty dx
\,\, (V(x)) ^2 + {\cal O} (p^{-5}).\label{3}\eeq Applying the
usual procedure of subtracting and adding the asymptotic behavior,
we rewrite the zeta function as \beq \zeta (s) =- \frac{\sin (\pi
s)} \pi \sum_\lambda \int\limits_{\sqrt{m^2 + \lambda^2}}^\infty
dp \left( p^2 - m^2 - \lambda ^2 \right) ^{-s} \frac \partial
{\partial p }\left\{ \ln s_{11} (ip) + \frac 1 {2p}
\int\limits_{-\infty} ^\infty dx \,\, V(x) \right\}
\nn\\+\frac{\sin (\pi s)} \pi \sum_\lambda \int\limits_{\sqrt{m^2
+ \lambda^2}}^\infty dp \left( p^2 - m^2 - \lambda ^2 \right)
^{-s} \frac
\partial {\partial p }\left(\frac 1 {2p} \int\limits_{-\infty} ^\infty
dx \,\, V(x) \right).\label{4}\eeq Here we have only subtracted
the first term in the asymptotic expansion, which turns out to be
sufficient in $d=2$ and $d=3$. Subtracting more terms might be
numerically helpful in these dimensions, and furthermore it is
necessary in $d\geq 4$. The $d=4$ result is given in section
\ref{4dcomp}. But for ease of presentation we first focus on
$d\leq 3$; the generalization is straightforward.

The above representation of the zeta function shows that
$$\zeta (s) =
\zeta _f (s) + \zeta _{as} (s)$$ with \beq \zeta _f (s) =-
\frac{\sin (\pi s)} \pi \sum_\lambda \int\limits_{\sqrt{m^2 +
\lambda^2}}^\infty dp \left( p^2 - m^2 - \lambda ^2 \right) ^{-s}
\frac
\partial {\partial p }\left\{ \ln s_{11} (ip) + \frac 1 {2p}
\int\limits_{-\infty} ^\infty dx \,\, V(x) \right\} \nn\eeq and
\beq \zeta _{as} (s) = - \frac{\sin \pi s} {2\pi}
\int\limits_{-\infty} ^\infty dx \,\, V(x) \sum_\lambda
\int\limits_{\sqrt{m^2 + \lambda^2} }^\infty dp \,\,\frac {\left(
p^2 - m^2 - \lambda^2 \right)^{-s}}{p^2}. \nn\eeq By construction,
$\zeta _f (s)$ is well behaved about $s=0$ and $\zeta _f ' (0)$ is
trivially calculated, \beq \zeta _f ' (0) = \sum _\lambda \left\{
\ln s_{11} \left( i \sqrt{m^2 + \lambda^2}\right) + \frac 1 {2
\sqrt{m^2 + \lambda^2}} \int\limits_{-\infty}^\infty dx \,\, V(x)
\right\} .\nn\eeq Cancellations between $\zeta_f ' (0)$ and
$\zeta_{as} ' (0)$ are expected to occur if we expand the above
expression further for $| \lambda| \gg 1$ because this has been
observed in \cite{dunn06-39-11915} for spherically symmetric
potentials. In the range $\lambda \gg 1$ we use
$$(m^2 + \lambda^2 ) ^ {-1/2} = \frac 1 {|\lambda|} \left( 1 +
{\cal O} \left( |\lambda|^{-3}\right)\right).$$ This allows us to
rewrite the answer for $\zeta_f ' (0)$ in the form \beq \zeta_f '
(0) &=& \sum_\lambda \left\{ \ln s_{11} \left( i \sqrt{m^2 +
\lambda^2}
\right) + \frac 1 {2 | \lambda|} \intv V(x) \right\} \nn\\
& & + \frac 1 2 \sum_\lambda \left\{ \frac 1 {\sqrt{m^2 +
\lambda^2}} - \frac 1 {|\lambda|}\right\} \,\, \intv V(x) .\nn\eeq
For $\zeta_{as} (s) $ we first perform the $p$-integration to obtain
\beq \zeta_{as} &=& -\frac{\sin \pi s}{2\pi} \frac{\Gamma (1-s)
\Gamma \left( \frac 1 2 + s\right)}{\sqrt{\pi}} \intv
V(x)\sum_\lambda (m^2 + \lambda^2 )^{-s-1/2} .\nn\eeq This makes
the introduction of the zeta function of the Laplace-type operator
in the $y$-coordinate necessary, and we define $$\zeta^{m}_y (s) =
\sum _\lambda (m^2 + \lambda^2)^{-s}.$$ Therefore, \beq \zeta
_{as} (s) &=&- \frac{\sin \pi s}{2\pi}\frac{\Gamma (1-s) \Gamma
\left( \frac 1 2 + s\right)}{\sqrt{\pi}} \zeta_y^{m} \left( s +
\frac 1 2 \right)
\,\, \intv V(x)\nn\\
&=& - \frac{\Gamma \left( s+\frac 1 2 \right)}{2 \sqrt \pi \Gamma
(s) } \zeta_y^m \left( s + \frac 1 2 \right) \intv V(x) .\nn \eeq
Applying a similar procedure as in $\zeta _f (s)$, we write \beq
\zeta _y ^m \left( s + \frac 1 2 \right) = \zeta _y ^0 \left( s +
\frac 1 2 \right) + \sum_\lambda \left\{ (m^2 + \lambda^2 )
^{-s-1/2} - | \lambda| ^{-2s-1}\right\} .\nn\eeq Using this
splitting, we continue \beq \zeta_{as} ' (0)& =& - \frac 1 2 \intv
V(x) \sum_\lambda \left\{ (m^2 + \lambda ^2 )^{-1/2} -
|\lambda|^{-1} \right\} \nn\\
& &- \left. \frac 1 {2 \sqrt \pi} \intv V(x) \,\, \frac d {ds}
\right|_{s=0} \left( \frac { \Gamma \left( s + \frac 1 2
\right)}{\Gamma (s) } \zeta _y ^0 \left( s + \frac 1 2 \right)
\right) .\nn\eeq In order to provide an answer as explicitly as
possible, which means an answer that will be applicable to
examples as easily as possible, we will assume that the structure
of the zeta function on $M$ is the standard one, i.e. for
$s\approx 0$ we assume \cite{seel68-10-288} \beq \zeta _y ^0
\left( s+\frac 1 2 \right) = \frac 1 s \mbox{Res } \zeta_y ^0
\left( \frac 1 2 \right) + PP \, \zeta_y ^0 \left( \frac 1 2
\right) + {\cal O} (s) .\nn\eeq This structure will apply for
example for a smooth potential $W(y)$ on a compact manifold $M$,
or for non-compact manifolds if the potential $W(y)$ is falling
off sufficiently fast at infinity. With this structure assumed, it
is easy to show that \beq \zeta _{as} ' (0) &=& - \frac 1 2 \intv
V(x) \left\{ \sum_\lambda \left[ (m^2 + \lambda^2 )^{-1/2} - |
\lambda |^{-1}\right] + PP \zeta _y ^0 \left( \frac 1 2 \right) -2
\ln 2 \,\,\mbox{Res }\zeta _y ^0 \left( \frac 1 2 \right) \right\}
.\nn\eeq Adding up the contributions from $\zeta_f ' (0)$ and
$\zeta _{as} ' (0)$ we find \beq \zeta ' (0) &=& \sum_\lambda
\left\{ \ln s_{11} \left( i \sqrt{ m^2 + \lambda^2}\right) + \frac
1 {2 | \lambda|} \intv V(x) \right\}\nn\\
& &+ \left\{ - \frac 1 2 PP \zeta _y ^0 \left( \frac 1 2 \right) +
\ln 2 \,\, \mbox{Res } \zeta _y ^0 \left( \frac 1 2 \right)
\right\}\int\limits_{-\infty} ^\infty dx \,\, V(x) .\label{5}\eeq
This seems to be the most compact answer in $d=2$ and $d=3$ one
can find. If there are $d_0$ eigenvalues $\lambda=0$, one needs to
add the contribution $d_0 \ln s_{11} (im)$ to the above answer,
with the understanding that the summation over $\lambda$ in the
subsequent quantities always omits the eigenvalue $\lambda =0$.

Once the manifold $M$ is specified to be a particular manifold,
thereby defining the zeta functions $\zeta_y(s)$ and $\zeta^0_y(s)$, the
above results can be made completely explicit. This is best seen
for cases like the torus and the sphere with $W(y) =0$ where final
answers are given in terms of well known special functions. In
principle it could also be done for the example of a ball, but the
associated zeta functions are not readily expressed in terms of
known functions, see \cite{bord96-37-895,bord96-182-371}, and
therefore we do not present details.

\subsection{Example of the torus}

{\bf d=2}: Let us assume one toroidally compactified dimension of
length $L$. Then $\lambda_n^2 = (2\pi n/ L)^2$, $n\in\intgs$, so
$d_0 = 1$, and \beq\zeta_y^0 (s) = \sum_{n\in \intgs / \{0\}}
\left( \frac{ 2\pi n}{L} \right)^{-2s} = 2 \left( \frac{2\pi} L
\right) ^{-2s} \zeta_R (2s) .\label{6n}\eeq In this case, \beq
\mbox{Res } \zeta_y ^0 \left( \frac 1 2 \right) = \frac L {2\pi},
\quad \quad \mbox{PP } \zeta_y ^0 \left( \frac 1 2 \right) = \frac
L \pi \left( \gamma - \ln \frac{2\pi } L \right) .\nn\eeq
Therefore, the final answer with one toroidal dimension reads \beq
\zeta ' (0) &=& \ln s_{11} ( im) + 2 \sum_{n=1}^\infty \left\{ \ln
s_{11} \left( i \left[ m^2 + \left( \frac{2\pi n}{L} \right)^2
\right] ^{1/2}
\right) + \frac L {4\pi n} \intv V(x) \right\} \nn\\
& &- \frac L {2\pi} \left( \gamma + \ln \frac L {4\pi } \right)
\intl_{-\infty}^\infty dx \,\, V(x) .\label{6}\eeq
{\bf d=3}: Let
us now assume two toroidally compactified dimensions of lengths
$L_1$ and $L_2$. In this case the zeta function associated with
the $y$-differential operator is an Epstein zeta function
\cite{epst03-56-615,epst07-63-205}, \beq \zeta _y ^0 (s) = (2\pi)
^{-2s} \sum_{(n,j)\in \intgs^2/\{0\}} \left[ \left( \frac n {L_1}
\right)^2 + \left( \frac j {L_2}\right)^2 \right]^{-s} .\nn\eeq
These functions, often defined as \beq Z_2 (s; w_1, w_2 )
=\sum_{(n,j)\in \intgs^2/\{0\}} \left[ w_1 n ^2 + w_2 j ^2
\right]^{-s} ,\nn\eeq have well understood analytical
continuations \cite{ambj83-147-1,eliz90-31-170,kirs94-35-459}.
Particularly suitable if one of the compactification lengths is
sent to zero is \beq Z_2 (s; w_1, w_2) &=& \frac 2 {w_2^s} \zeta_R
(2s) + \frac {2 \sqrt \pi} {\sqrt {w_2} } w_1 ^{\frac 1 2 -s}
\zeta_R (2s-1) \frac{\Gamma
\left( s- \frac 1 2 \right)} { \Gamma (s)} \nn\\
& &+ \frac {8 \pi^s}{\Gamma (s) \sqrt {w_2}} \sum_{n=1}^\infty
\sum_{j=1}^\infty \left[\sqrt{w_1 w_2} \frac j n\right]^{\frac 1 2
-s} K_{\frac 1 2 -s} \left( 2\pi
nj\sqrt{\frac{w_1}{w_2}}\right).\label{7}\eeq From this analytical
continuation one easily derives that $\mbox{Res }\zeta_y^0
(1/2)=0$ (the singular contributions from the first two terms
cancel), and \beq \zeta _y ^0 \left( \frac 1 2 \right) &=& \frac
{L_2} \pi \left( \gamma + \ln \left[ \frac 1 {4\pi } \frac {L_2} {
L_1}\right] \right)+ \frac{ 4 L_2} \pi \sum_{n=1}^\infty
\sum_{j=1}^\infty K_0 \left( 2\pi n j \frac {L_2} {L_1} \right)
.\nn\eeq Using this in (\ref{5}), we find \beq \zeta ' (0) &=& \ln
s_{11} (im) \nn\\
& &-  \left\{ \frac{L_2} {2\pi} \left( \gamma + \ln \left[ \frac
 1 {4\pi } \frac{L_2} {L_1}\right]\right) + \frac{ 2L_2} \pi
 \sum_{n=1}^\infty \sum_{j=1}^\infty K_0 \left( 2\pi n j \frac
 {L_2}{L_1}\right) \right\} \intv V(x) . \nn\\
& &+ \sum_{(n,j)\in \intgs^2/\{0\}} \left\{ \ln s_{11} \left( i
\sqrt{ m^2 + \left( \frac {2\pi n} {L_1}\right)^2 + \left( \frac{
2\pi j}{L_2}\right)^2 }\right)\right. \nn\\
& &\hspace{5.0cm}\left.+ \frac 1 {4\pi} \frac 1 {\sqrt{ \left(
\frac n {L_1}\right)^2 + \left( \frac j {L_2}\right)^2 } } \intv
V(x) \right\} \label{8}\eeq

In case we consider toroidal compactification of equal sides, the
results look even simpler. Due to results of Hardy
\cite{hard19-49-85}, with $$\beta (s) = \sum_{n=0} ^\infty (-1)^n
(2n+1)^{-s},$$ in this case we have
$$\zeta _y ^0 (s) = \left( \frac L {2\pi} \right)^{2s} 4 \zeta _R
(s) \beta (s)
$$ and the final answer takes the simple form \beq \zeta ' (0)&=&
\ln s_{11} (im)- \frac L \pi \zeta _R \left( \frac 1 2 \right)
\beta \left( \frac 1 2 \right) \intl_{-\infty}^\infty dx \,\,
V(x).\label{9}\\
& & + \sum_{(n,j)\in \intgs^2/\{0\}} \left\{ \ln s_{11} \left( i
\sqrt{ m^2 + \left( \frac {2 \pi } L \right)^2 \left[ n^2 +
j^2\right]} \right) + \frac L {4\pi} \frac 1 {\sqrt {n^2 + j^2}}
\intv V(x) \right\} \nn \eeq
\subsection{Example of a sphere}
For $M$ a $d-1$ dimensional unit sphere the eigenvalues are known
to be \cite{erde55b} $$\lambda_l = l (l+d-2) = \left( l+\frac{d-2}
2 \right)^2 - \frac{(d-2)^2} 4 , \quad l \in \nats_0.$$ In case a
sphere of radius $a$ is considered the eigenvalues simply scale
like $1/a^2$. Therefore we will always assume $a=1$ as $a\neq 1$
follows trivially. The degeneracy ${\rm deg} (l;d-1)$ for each
eigenvalue is given by $${\rm deg} (l;d-1) = (2l+d-2)
\frac{(l+d-3)!} { l! (d-2)!}, \quad l \in \nats_0.$$ To keep the
analysis as simple as possible, we assume $W(y) = (d-2)^2/4$,
which corresponds to conformal coupling in $(d-1)$ dimensions. The
eigenvalues then become a complete square,
$$\lambda_l = \left( l + \frac{d-2} 2 \right)^2.$$ Remarks about
the case where $W(y)$ is an arbitrary constant are made in the
Appendix
\ref{app-sphere}. \\
{\bf d=2}: This is identical to the torus case with $L=2\pi$ and
nothing more needs to be said.\\
{\bf d=3}: On the 2 sphere we have $\lambda_l = \left( l + \frac 1
2 \right)^2$ with degeneracy ${\rm deg} (l;d-1) = 2l+1$. Therefore
we see \beq\zeta_y^0 (s) = 2 \zeta_H \left(2s-1; \frac 1 2 \right)
= 2 (2^{2s-1} -1 ) \zeta_R (2s-1). \label{zetathree}\eeq Here
$\zeta_H (s;b)$ denotes as usual the Hurwitz zeta function. The
relevant properties of $\zeta_y^0 (s)$ are
$$\mbox{Res } \zeta_y ^0 \left( \frac 1 2 \right) = 0 , \quad
\quad \mbox{PP } \zeta _y^0 \left( \frac 1 2 \right) = 0.$$ This
shows the final answer for eq. (\ref{5}) reads \beq \zeta' (0) =
\sum_{l=0}^\infty \left\{ (2l+1) \ln s_{11} \left( i \sqrt{m^2 +
\left( l+\frac 1 2 \right)^2} \right) +
\int\limits_{-\infty}^\infty dx \,\, V(x) \right\} .\nn\eeq

\subsection{Zeta function construction with over-subtraction}
\label{4dcomp}

In $d=2$ and $d=3$, for numerical convenience one might decide to
add and subtract more terms in eq. (\ref{4}) than just the leading
term from eq. (\ref{3}). Writing down the answer for the case with
$\lambda^2
> 0$, having in mind the needed changes if $\lambda =0$ occurs, we
get for the general case subtracting the two terms given in
(\ref{3}) \beq \zeta ' (0) &=& \sum_\lambda \left\{ \ln s_{11} (i
\sqrt{ m^2 + \lambda^2}) + \frac 1 {2 | \lambda|} \intv V(x) -
\frac 1 {8 |
\lambda|^3 } \intv V(x) ( V(x) + 2 m^2 ) \right\} \nn\\
&+& \left(\mbox{Res } \zeta _y ^0 \left( \frac 1 2 \right) \ln 2 -
\frac 1 2 PP \zeta _y ^0 \left( \frac 1 2 \right) \right) \intv
V(x) \nn\\
&+& \frac 1 8 \left( \mbox{PP }\zeta _y ^0 \left( \frac 3 2
\right) - \mbox{Res } \zeta_y^0 \left( \frac 3 2 \right) [-2 + \ln
4]\right) \intv V(x) ( V(x) + 2 m^2 ) .\label{10} \eeq This result
is also valid in $d=4$. For particular cases like the torus or the
sphere the final answer is easily found from known properties of
Epstein type zeta functions; some details are given in Appendices
\ref{app-torus} and \ref{app-sphere}. Higher dimensions and more
subtractions could be considered if necessary.

\section{Conclusions}

To conclude, we have presented new simplified explicit formulas for the one-loop vacuum energy when the underlying spacetime manifold is separable. The results have been derived using the zeta function  method in conjunction with dimensional regularization and renormalization. The relation to dimensionally regularized Feynman diagrams is exactly as discussed in \cite{dunn06-39-11915}. The cases considered here include (i) a static, radially symmetric background field in $(3+1)$- and $(2+1)$-dimensions, at both zero and nonzero temperature; and (ii) a nontrivial domain-wall profile with a compact transverse manifold such as a sphere or a torus. The analysis is ultimately based on the Gel'fand-Yaglom theorem for the determinant of an ordinary differential operator, but extended to incorporate the necessary regularization and renormalization that appears in higher dimensions. Computationally, the final expressions are finite and convergent, and are considerably simpler than the corresponding expressions for example  in \cite{bord00-61-085008,bord96-53-5753}.

\section*{Acknowledgments} GD thanks the US DOE for support
through grant DE-FG02-92ER40716.  KK acknowledges support by the
NSF through grant PHY-0757791.

\appendix

\section{Asymptotic decomposition of $\zeta_{as} (s)$}
In this appendix we provide some details for the analysis of
$\zeta_{as} (s)$ and of the finite temperature case for the
situation of a spherically symmetric background field, see Section
II.

Eqs. (\ref{zeta-as}), (\ref{zeta-f-split}), (\ref{zeta-f-1}) and
(\ref{zeta-f-2}) suggest to introduce the function  \beq f(s,c,b) = \frac 1
\beta  \snmp \sln {\rm deg}(l; d-1) \nu^{-2s-c} \pmr
^{-s-b}.\nn\eeq We then find the representation \beq \zeta_{as}
(s) &=& - \frac{\Gamma \left( s + \frac 1 2 \right)}{2 \sqrt \pi
\Gamma (s)} \intl_0^\infty dr \,\, W(r) r^{1+2s} f\left( s,1,\frac
1 2\right) \label{zasstart}\\
& &+ \frac{\Gamma \left( s + \frac 3 2 \right)}{4 \sqrt \pi \Gamma
(s)} \intl_0^\infty dr \,\, W(r) r^{1+2s} \left( W(r) r^2 - \frac
1 2 \right) f \left( s, 3, \frac 3 2 \right) \nn\\
& &+ \frac{\Gamma \left( s + \frac 5 2 \right)} {2 \sqrt \pi
\Gamma (s) } \intl_0^\infty dr \,\, W(r) r^{1+2s} f \left( s , 3 ,
\frac 5 2 \right) \nn\\
& &- \frac{\Gamma \left( s + \frac 7 2 \right)} { 6 \sqrt \pi
\Gamma (s) } \intl_0^\infty dr \,\, W(r) r^{1+2s} f \left( s , 3 ,
\frac 7 2 \right) .\nn\eeq In order to evaluate $\zeta _{as} '
(0)$ we mimic the process employed for $\zeta _f ' (0)$. We
explain the details for $f(s,1,1/2)$ as all other terms are
obtained accordingly. First note that \beq f\left( s,1,\frac 1 2
\right) &=& \frac 1 \beta \snmp \sln {\rm deg}(l; d-1) (\nu^2 +
p_n^2 r^2) ^{-s-\frac 1 2}\times\nn\\
& &\hspace{2cm} \left[ \pmrr^{-s-\frac 1 2} - 1 + \left( s+\frac 1
2 \right)
\frac{m^2 r^2}{\nu^2 + p_n^2 r^2} \right] \nn\\
& &+ \frac 1 \beta \snmp \sln {\rm deg}(l; d-1) (\nu^2 + p_n^2
r^2)^{-s-\frac 1 2} \left( 1 - \left( s + \frac 1 2 \right)
\frac{m^2 r^2}{\nu^2 + p_n^2 r^2} \right).\nn\eeq The first term
is by construction well defined at $s=0$, and for the second term an
analytical continuation to $s=0$ has to be performed. To this aim
we introduce the Epstein type zeta function \beq E_d^{(k)} (s,a)
&=& \frac 1 \beta \snmp \sln {\rm deg}(l; d-1) \frac{\nu^k}{(\nu^2
+ a^2 n^2)^s} .\label{epty}\eeq We then have for the first term in
(\ref{zasstart}) $$\zeta_{as,1} (s) = \zeta_{as,1} ^{(1)} (s) +
\zeta_{as,1}^{(2)} (s)$$ with \beq \zeta_{as,1}^{(1)} (s) &=& -
\frac 1 {2 \sqrt \pi} \frac{\Gamma \left( s + \frac 1 2
\right)}{\Gamma (s ) }
\intl_0^\infty dr \,\, W(r) r^{1+2s}\times \nn\\
& &\hspace{2.0cm}\left\{ E_d^{(0)} \left( s + \frac 1 2 , \frac{
2\pi r} \beta \right) - \left( s + \frac 1 2 \right) m^2 r^2
E_d^{(0)} \left( s + \frac 3 2 , \frac{2\pi r } \beta
\right)\right\},\nn\\
\zeta_{as,1}^{(2)} (s) &=& - \frac 1 2 \intl_0^\infty dr \,\, W(r)
r \,\,\frac 1 \beta \snmp \sln {\rm deg}(l; d-1) (\nu^2 + p_n^2
r^2 )
^{-s-\frac 1 2}\times\nn\\
& &\hspace{2cm} \left[ \pmrr ^{-s-\frac 1 2} - 1 + \left(s+ \frac
1 2 \right) \frac{m^2 r^2} {\nu^2 + p_n^2 r^2} \right].\nn\eeq
Proceeding in the same fashion with the other terms in
(\ref{zasstart}), equations (\ref{zeta-as-1}) and
(\ref{zeta-as-2}) are established. Whereas the contribution
$\zeta_{as}^{(2)\prime} (0)$ cancels with terms in $\zeta_f '
(0)$, namely with the last term in (\ref{zeta-prime-f-new}), the
term $\zeta_{as}^{(1)} (s)$ contributes \beq
\zeta_{as}^{(1)\prime} (0)&=&
\left. - \frac 1 {2\sqrt \pi} \frac d {ds} \right|_{s=0}
\frac{\Gamma \left( s + \frac 1 2 \right)} {\Gamma (s)}
\intl_0^\infty dr \,\, W(r) r^{1+2s} \times\nn\\
& &\hspace{2.0cm}\left( E_d^{(0)} \left( s + \frac 1 2 ,
\frac{2\pi r} \beta\right) - \left( s + \frac 1 2 \right) m^2 r^2
E_d^{(0)} \left( s + \frac 3 2 , \frac{2\pi r}
\beta \right)\right) \nn\\
& &\left.+\frac 1 {4 \sqrt \pi} \frac d {ds} \right|_{s=0}
\frac{\Gamma \left( s + \frac 3 2 \right)} {\Gamma (s)}
\intl_0^\infty dr \,\, W(r) r^{1+2s} \left(W(r) r^2 - \frac 1 2
\right) E_d^{(0)} \left( s + \frac 3 2 , \frac{2\pi r}
\beta\right)  \nn\\
& &\left.+\frac 1 {2 \sqrt \pi} \frac d {ds} \right|_{s=0}
\frac{\Gamma \left( s + \frac 5 2 \right)} {\Gamma (s)}
\intl_0^\infty dr \,\, W(r) r^{1+2s} E_d^{(2)} \left( s + \frac 52
, \frac{2\pi r}
\beta\right)  \nn\\
& &\left. -\frac 1 {6 \sqrt \pi} \frac d {ds} \right|_{s=0}
\frac{\Gamma \left( s + \frac 7 2 \right)} {\Gamma (s)}
\intl_0^\infty dr \,\, W(r) r^{1+2s}  E_d^{(4)} \left( s + \frac 7
2 , \frac{2\pi r} \beta\right).  \nn\eeq To find an explicit
answer for $\zeta_{as}^{(1)\prime} (0)$ we are left to do the
analysis of the zeta function $E_d^{(k)} (s,a)$ in eq.
(\ref{epty}). To perform the limit $\beta \to \infty$ the best way
to proceed is to perform a Poisson resummation in the
$n$-summation \cite{ambj83-147-1,eliz90-31-170,kirs93-26-2421}.
This leads to the following result, \beq E_d ^{(k)} (s,a) &=&
\frac{ \sqrt \pi} {a \beta} \frac{\Gamma \left( s- \frac
1 2\right)} {\Gamma (s)} \sln {\rm deg}(l; d-1) \nu^{k+1 -2s} \label{relzeta}\\
& &+ \frac{ 4 \pi ^s}{\beta \Gamma (s) a^{s+\frac 1 2}} \sln {\rm
deg}(l; d-1) \nu^{k+\frac 1 2 -s} \sum_{n=1}^\infty n^{s-\frac 1 2
} K_{\frac 1 2 -s } \left( 2 \pi \nu \frac n a \right).\nn\eeq The
second line is analytic for all values of $s$, the first line
contains first order singularities at certain $s$-values that need
to be considered in detail. Using the Laurent series of
meromorphic functions with a first order pole at $s=s_0$ we expand
$$E_d ^{(k)} (s,a) = \frac 1 {s-s_0} \mbox{ Res }E_d^{(k)} \left(
s_0,a\right) + \mbox{PP } E_d^{(k)} (s_0,a) + {\cal O} (s-s_0) ,
$$ to obtain $\zeta_{as}^{(1)\prime}(0)$ in the form \beq \zeta_{as}^{(1)\prime} (0)
& &- \frac 1 {16} \intl_0^\infty dr \,\, r W(r) \left[ 8 \left(
\mbox{PP } E_d^{(0)} \left( \frac 1 2 , \frac{ 2\pi r} \beta
\right) + \mbox{Res } E_d^{(0)} \left( \frac 1 2 , \frac{ 2\pi r}
\beta \right) \ln \left( \frac {r^2} 4 \right) \right) \right.
\nn\\
& &+(1-2 r^2 [ W(r) + 2 m^2]) \left( \mbox{PP } E_d ^{(0)} \left(
\frac 3 2 , \frac{ 2\pi r} \beta \right) + \mbox{Res }E_d^{(0)}
\left( \frac 3 2 , \frac{2\pi r} \beta \right) \left[ \ln \left(
\frac{r^2} 4 \right) +2 \right] \right) \nn\\
& &-6 \left( \mbox{PP }E_d ^{(2)} \left( \frac 5 2 , \frac{2\pi r}
\beta \right) + \mbox{Res } E_d ^{(2)} \left( \frac 5 2 , \frac{
2\pi r } \beta \right) \left[ \ln \left( \frac{ r^2} 4\right) +
\frac 8 3
\right] \right) \nn\\
& &\left. + 5 \left( \mbox{PP } E_d ^{(4)} \left( \frac 7 2 ,
\frac{2\pi r} \beta \right) + \mbox{Res } E_d ^{(4)} \left( \frac
7 2 , \frac{ 2\pi r} \beta \right) \left[ \ln \left( \frac{r^2} 4
\right) + \frac{ 46} {15} \right] \right)
\right].\label{zetapg}\eeq The relevant properties of $E_d^{(k)}
(s,2\pi r/\beta) $ are easily found from (\ref{relzeta}). First
note that the series over the Bessel functions $K_{1/2-s} (\nu n
\beta /r)$ vanishes exponentially fast as $\beta \to \infty$. Its
contribution at finite temperature is obtained by simply
substituting the $s$-values needed in (\ref{zetapg}). In some
detail, if we define $$C_d^{(k)} (s,a) =\frac{ 4 \pi ^s}{\beta
\Gamma (s) a^{s+\frac 1 2}} \sln {\rm deg}(l; d-1) \nu^{k+\frac 1
2 -s} \sum_{n=1}^\infty n^{s-\frac 1 2 } K_{\frac 1 2 -s } \left(
2 \pi \nu \frac n a \right),$$ then its contribution follows from
replacing $\mbox{PP }E_d^{(k)} (s, 2\pi r/\beta )$ by $C_d^{(k)}
(s, 2\pi r \beta )$ in (\ref{zetapg}); note there are no
contributions to the residue terms in (\ref{zetapg}). The
contributions so obtained have to be added to the zero temperature
result (\ref{4d-answer}) following from the first line in
(\ref{relzeta}).

So in the following, let us concentrate on the contributions from
the first line in (\ref{relzeta}), which are the $T=0$
contributions. In $d=4$ we have ${\rm deg} (l;d-1) = (2l+1)$ and
$\nu = l+1/2$ which shows \beq E_4^{(k)} \left(s,\frac{2\pi r}
\beta\right) = \frac 1 {\sqrt \pi r} \frac{\Gamma \left( s-\frac 1
2 \right)}{\Gamma (s)} \zeta_H \left(2s-k-2 ; \frac 1 2 \right).
\label{epsind4} \eeq The relevant expansions are \beq E_4 ^{(0)}
\left( s , \frac{2\pi r} \beta \right) &=& \frac 1 {24 \pi r}
\,\frac 1 {s-\frac 1 2} - \frac 1 {\pi r} \zeta_R ' (-1) + {\cal
O}
\left( s-\frac 1 2\right), \nn\\
E_4^{(0)} \left( s , \frac {2\pi r} \beta\right) &=& \frac 1 {\pi
s} \,\frac 1 {s-\frac 3 2 } + \frac{ 2} {\pi r} (-1
+\gamma + \ln 8 ) + {\cal O} \left( s- \frac 3 2 \right) , \nn\\
E_4^{(2)} \left( s , \frac{ 2\pi r} \beta \right) &=& \frac{2}
{3\pi r}\, \frac 1 {s-\frac 5 2 } + \frac{2} {9\pi r} ( -5 +
6\gamma + 18 \ln 2 ) + {\cal O} \left( s-\frac 5
2\right) , \nn\\
E_4 ^{(4)} \left( s , \frac{2\pi r} \beta \right) &=& \frac{8}{15
\pi r} \,\frac 1 {s-\frac 7 2 } + \frac{4} {15 \pi r} \left( -
\frac{47} {15} + 4 \gamma + 12 \ln 2\right) + {\cal O} \left(
s-\frac 7 2\right).\nn\eeq Using these in (\ref{zetapg}) the
result is eq. (\ref{zeta-as-1-final}), namely \beq
\zeta_{as}^{(1)\prime} (0) &=& \frac 1 {24 \pi}
\int\limits_0^\infty dr \,\, W(r) \left(12 \zeta_R ' (-1)+\gamma +
3 \ln 2 +6r^2 (W(r) + 2m^2) ( \gamma +\ln (4m r))\right)
.\nonumber\eeq As mentioned before, in $d=3$, eq.
(\ref{jost-asymptotic}) can only be used for $l\geq 1$. The
contributions from those modes are still given by (\ref{zetapg})
with the sum in (\ref{relzeta}) starting at $l=1$. For $l\geq 1$,
the degeneracy ${\rm deg} (l;2) =2$ and (\ref{relzeta}) is
replaced by \beq E_3^{(k)} (s,a) &=& \frac{2 \sqrt \pi} {a\beta}
\frac{\Gamma \left( s- \frac 1 2\right)} {\Gamma (s)}
\zeta _R (2s-k-1) \nn\\
& &+\frac{8 \pi^s}{\beta \Gamma (s) a^{s+\frac 1 2} }
\sum_{l=1}^\infty \sum_{n=1}^\infty l^{k+\frac 1 2 -s} n^{s-\frac
1 2 } K_{\frac 1 2 -s} \left( \frac{2\pi l n}
a\right).\label{relzed3}\eeq For reasons explained, concentrating
on $T=0$, the relevant expansions are \beq E_3^{(0)} \left( s,
\frac{2\pi r} \beta \right) &=& - \frac 1 {2\pi r}\, \frac 1
{s-\frac 1 2} - \frac{1} {\pi r} \ln (4\pi) + {\cal O}
\left( s- \frac 1 2 \right), \nn\\
E_3^{(0)} \left( s, \frac{2\pi r} \beta \right) &=& \frac{
\pi } {3r} + {\cal O} \left( s- \frac 3 2 \right) ,\nn\\
E_3^{(2)} \left( s , \frac{2\pi r} \beta\right) &=& \frac{2
\pi} {9r} + {\cal O} \left( s-\frac 5 2 \right) , \nn\\
E_3^{(4)} \left( s , \frac{2\pi r} \beta\right) &=& \frac{8 \pi}
{45r} + {\cal O} \left( s-\frac 7 2 \right) . \nn\eeq The $l=0$
contribution is \beq \zeta_0 (s) &=& \frac{\sin \pi s} {\pi\beta}
\snmp \intl_{\sqrt {\pmq}} ^\infty dk \,\, \left( \kpm
\right)^{-s} \frac d {dk} \ln f_0 (ik) .\nn\eeq Adding and
subtracting the large-$k$ behavior of $\ln f_0 (ik)$ this is
rewritten along the lines presented as \beq \zeta_0 (s) &=&
\frac{\sin \pi s} {\pi\beta} \snmp \intl_{\sqrt {\pmq}} ^\infty dk
\,\, \left( \kpm \right)^{-s} \frac d {dk} \left[ \ln f_0 (ik) -
\frac
1 {2k} \intl_0^\infty dr \,\, W(r) \right]\nn\\
& &\left.-\frac 1 {2\sqrt \pi} \frac{\Gamma \left( s+\frac 1 2
\right)} {\Gamma (s)} Z \left( s+\frac 1 2 , \left( \frac{2\pi}
\beta\right)^2\right|m^2\right) \intl_0^\infty dr \,\, W(r),
\nn\eeq where \cite{ambj83-147-1,eliz90-31-170,kirs93-26-2421}
\beq Z \left( s,a^2|m^2\right)&=& \frac 1 \beta \sum_{n=-\infty}
^\infty \left( a^2 n^2 +m^2\right)^{-s} \nn\\
&=& \frac{\sqrt \pi} {a\beta} \frac{\Gamma \left( s- \frac 1 2
\right)}{\Gamma (s)} m^{1-2s} + \frac{4\pi ^s m^{\frac 1 2
-s}}{\beta \Gamma (s) a} \sum_{n=1}^\infty \left(\frac n
a\right)^{s-\frac 1 2} K_{\frac 1 2 -s} \left( 2\pi m \frac n
a\right).\label{epnull}\eeq At $T=0$ this gives for the
determinant of the $l=0$ contribution \beq \zeta_0'(0) &=& - \frac
1 {2\pi} \intl_{-\infty} ^\infty dp \left[ \ln f_0 (i \sqrt{p^2
+m^2}) - \frac 1 {2\sqrt{p^2 +m^2}} \intl_0^\infty dr \,\, W(r)
\right] + \frac 1 {2\pi} \ln m \intl_0^\infty dr \,\, W(r)
.\nn\eeq At finite temperature a series over Bessel functions
resulting in (\ref{epnull}) has to be added as described below
(\ref{zetapg}).

Adding up the $l=0$ and $l\geq 1$ contributions, we obtain \beq
\zeta ' (0) &=& - \frac 1 {2\pi} \intl_{-\infty} ^\infty dp \left[
\ln f_0 (i\sqrt{p^2+m^2}) - \frac 1 {2\sqrt{p^2+m^2}}
\intl_0^\infty dr \,\, W(r) \right] \nn\\
& &-\frac 1 \pi \sum_{l=1}^\infty \intl_{-\infty}^\infty dp \left[
\ln f_l (i\sqrt{p^2+m^2}) - \frac 1 2 \intl_0^\infty dr\,\,
\frac{rW(r)} {(l^2 + p^2r^2)^{\frac 1 2}}\right.\nn\\
& &+\frac 1 8 \intl_0^\infty dr \,\, \frac{r^3 W(r) (W(r) +
2m^2)}{(l^2 + p^2 r^2)^{\frac 3 2}}\nn\\
& &\left. -\frac 1 {16} \intl_0^\infty dr \,\, \frac{rW(r)} {(l^2
+ p^2 r^2)^{\frac 3 2}} \left( 1-\frac {6l^2}{l^2 + p^2 r^2} +
\frac{5l^4}{(l^2 + p^2 r^2)^2}\right)\right]\nn\\
& &+\frac 1 {16\pi} \intl_0^\infty dr \,\, W(r) \left[
\frac{\pi^2} 9 (1+6r^2 (W(r) + 2m^2) ) + 8 \ln (2\pi
mr)\right].\nn\eeq This final answer involves oversubtractions we
made, which, despite the fact that it makes a numerical evaluation
easier has the disadvantage of looking more complicated.
Subtracting what is strictly necessary for the procedure in $d=3$
we obtain \beq \zeta ' (0) &=&- \frac 1 {2\pi} \intl_{-\infty}
^\infty dp \left[ \ln f_0 (i\sqrt{p^2+m^2}) - \frac 1
{2\sqrt{p^2+m^2}}
\intl_0^\infty dr \,\, W(r) \right] \nn\\
& &-\frac 1 \pi \sum_{l=1}^\infty \intl_{-\infty}^\infty dp \left[
\ln f_l (i\sqrt{p^2+m^2}) - \frac 1 2 \intl_0^\infty dr\,\,
\frac{rW(r)} {(l^2 + p^2r^2)^{\frac 1 2}}\right]\nn\\
& &+\frac 1 {2\pi} \intl_0^\infty dr \,\, W(r) \ln (2\pi
rm).\nn\eeq

\section{Zeta function properties on the torus}
\label{app-torus}

In this appendix we collect the information needed to evaluate the
expression for $\zeta ' (0)$ given in eq. (\ref{10}) for the
example of a torus.

To exploit eq. (\ref{10}) for the torus, in $d=2$ the only
additional information needed beyond eq. (\ref{6}) is $\zeta _y ^0
(3/2)$. From eq. (\ref{6n}) $$\mbox{Res } \zeta_y^0 \left( \frac 3
2 \right) =0 \quad \mbox{and}\quad \zeta_y^0 \left( \frac 3 2
\right) = \frac{L^3} {4\pi^3} \zeta_R (3) .$$ In $d=3$ we use eq.
(\ref{7}) to see $\mbox{Res }\zeta_y^0 (3/2)=0$ and \beq \zeta_y
^0 \left( \frac 3 2 \right) &=& Z_2 \left(
\frac 3 2 ; \left(\frac{ 2\pi} {L_1}\right)^2 , \left(\frac{2\pi} {L_2}\right)^2 \right) \nn\\
&=& \frac {L_2^3} {4\pi^3} \zeta_R (3) + \frac{L_1^2 L_2} {12 \pi}
+ \frac{ 2L_1 L_2^2}{\pi^2} \sum_{n=1}^\infty \sum_{j=1}^\infty
\frac n j K_1 \left( 2\pi nj \frac {L_2}{L_1}\right).\nn\eeq For
$d=4$ we introduce \beq Z_3 (s; w_1,w_2,w_3) = \sum_{(n_1,n_2,n_3)
\in \intgs^3 \{ 0 \} } (w_1 n_1^2+w_2 n_2^2 + w_3 n_3^2)^{-s}
\nn\eeq and have $$\zeta _y ^0 (s) = Z_3 \left( s;
\left(\frac{2\pi} {L_1}\right)^2 , \left(\frac{2\pi } {L_2}
\right)^2, \left(\frac{2\pi } {L_3}\right)^2 \right).$$ The
analytical continuation of $Z_3 (s; w_1,w_2,w_3)$ to a meromorphic
function in the complex plane reads
\cite{ambj83-147-1,eliz90-31-170,kirs93-26-2421} \beq Z_3 (s;
w_1,w_2,w_3) &=& \frac 2 {w_3^s} \zeta _R (2s) + \sqrt {
\frac{\pi} {w_3}}\,\, \frac{\Gamma \left( s-\frac 1 2
\right)}{\Gamma (s)} Z_2 \left( s-\frac 1 2 ; w_1, w_2\right)
\label{epst}\\ & & \hspace{-3.0cm}+\frac{4\pi^s} {\Gamma (s) \sqrt
{w_3}}\sum_{n=1}^\infty \sum_{(n_1,n_2) \in \intgs^2 / \{0\}}
\left( \frac { \left[ w_1 n_1^2 + w_2 n_2^2\right] w_3 }{n^2}
\right) ^{\frac 1 2 \left( \frac 1 2 - s \right)} K_{\frac 1 2 -s}
\left( \frac{ 2\pi n } {\sqrt{w_3}} \sqrt{ w_1 n_1^2 + w_2 n_2 ^2
} \right) , \nn\eeq which we need to analyze further about the
points $s= 1/2$ and $s=3/2$. As intermediate results we note that
\beq \mbox{Res } Z_3 \left( \frac 1 2 ; w_1 , w_2 , w_3 \right)
&=& \frac 1 {\sqrt{w_3}} + \frac{Z_2 ( 0; w_1,w_2)}{\sqrt{w_3}} ,
\nn\\
\mbox{PP } Z_3 \left( \frac 1 2 ; w_1 , w_2 , w_3 \right) &=&
\frac{ 2\gamma - \ln w_3 }{\sqrt {w_3}} + \frac 1 {\sqrt{ w_3}}
\left[ Z_2 ' (0; w_1,w_2) + Z_2 ( 0 ; w_1, w_2) \ln 4 \right]
\nn\\
& &+ \frac 4 {\sqrt{w_3}} \sum_{n=1} ^\infty \sum_{(n_1,n_2) \in
\intgs^2 / \{ 0 \}}  K_0 \left( \frac{ 2\pi n } {\sqrt{
w_3}} \sqrt{ w_1 n_1^2 + w_2 n_2 ^2 } \right) , \nn\\
\mbox{Res } Z_3 \left( \frac 3 2 ; w_1, w_2, w_3\right) &=& \frac
2 {\sqrt{w_3}} \mbox{Res } Z_2 ( 1; w_1 , w_2 ) , \nn\\
\mbox{PP } Z_3 \left( \frac 3 2 ; w_1,w_2,w_3\right) &=& \frac 2
{w_3 ^{\frac 3 2 }} \zeta_R (3) + \frac 2 {\sqrt{ w_3}} \left[
\mbox{PP } Z_2 (1; w_1,w_2) + \mbox{Res } Z_2 (1; w_1,w_2) ( \ln 4
-2)\right] \nn\\
& &+\frac{8 \pi } {w_3} \sum_{n=1}^\infty \sum_{(n_1,n_2) \in
\intgs ^2 / \{ 0 \}} \frac n {\sqrt{ w_1 n_1^2 + w_2 n_2 ^2 }} K_1
\left( \frac{2\pi n} {\sqrt{w_3}} \sqrt{ w_1 n_1^2 + w_2 n_2^2 }
\right) .\nn\eeq Eq. (\ref{7}) is then used to find the relevant
properties of $Z_2 (s; w_1,w_2)$ about $s=0$ and $s=1$,
namely \beq Z_2 (0; w_1, w_2) &=& -1, \nn\\
Z_2 ' (0; w_1,w_2) &=& \ln \left(\frac{ w_2} { 4\pi^2}\right) +
\frac \pi 3 \sqrt{ \frac{w_1} { w_2}} - 4 \sum_{n=1}^\infty \ln
\left(
1-e^{-2\pi n \sqrt{ \frac { w_1} { w_2}}}\right),\nn\\
\mbox{Res } Z_2 (1;w_1,w_2) &=& \frac \pi {\sqrt{w_1 w_2}} , \nn\\
\mbox{PP } Z_2 (1;w_1,w_2) &=& \frac{\pi^2} {3w_2} + \frac \pi
{\sqrt{w_1 w_2}} ( 2 \gamma - \ln (4 w_1)) - \frac{ 4\pi} {(w_1
w_2)^{\frac 3 4 }} \sum_{n=1}^\infty \ln \left( 1-e^{-2\pi n
\sqrt{\frac {w_1}{w_2}}}\right).\nn\eeq These results are used to
produce the final answer for eq. (\ref{10}) in $d=4$ by
substituting \beq \mbox{Res } Z_3 \left( \frac 1 2 ; \we , \wz,
\wdd \right) &=& 0 ,
\nn\\
\mbox{PP }Z_3 \left( \frac 1 2 ; \we ,\wz , \wdd \right) &=&\nn\\
& &\hspace{-3.0cm} \frac{L_3}{ 2 \pi } \left[ 2 \gamma + \ln
\left( \frac{L_3}{16 \pi ^2 L_2 } \right) + \frac \pi 3 \frac{L_2}
{L_1} - 4 \sum_{n=1}^\infty \ln \left( 1-e^{- 2\pi n
\frac{L_2}{L_1}}\right)
\right] \nn\\
& &\hspace{-3.0cm}+ \frac{2L_3} \pi \sum_{n=1}^\infty \sum_{(n_1,
n_2)\in \intgs ^2 / \{ 0 \}} K_0 \left( 2\pi n L_3 \sqrt{
\frac{n_1^2}{L_1^2} + \frac{n_2^2}{L_2^2}} \right),\nn\\
\mbox{Res }Z_3 \left( \frac 3 2 ; \we, \wz, \wdd \right) &=& \frac
1
{4\pi^2} L_1 L_2 L_3 , \nn\\
\mbox{PP } Z_3 \left( \frac 3 2 ; \we , \wz , \wdd \right) &=&
\frac 1 {4 \pi^3} L_3 ^3 \zeta_R (3)\nn\\
& & \hspace{-5.0cm}+ \frac{L_3} \pi \left[ \frac{L_2^2} {12} +
\frac{ L_1 L_2 } {4\pi} \left( 2 \gamma - 2 - 2 \ln \left( \frac{
2\pi } {L_1} \right)\right) - \frac{ (L_1 L_2)^{\frac  3 2
}}{2\pi^2} \sum_{n=1}^\infty\ln \left( 1-
e^{-2\pi n \frac{L_2}{L_1} }\right) \right] \nn\\
& &\hspace{-5.0cm}+\frac {L_3^2} {\pi^2}  \sum_{n=1}^\infty
\sum_{(n_1,n_2)\in \intgs^2 / \{ 0 \} }\frac n
{\sqrt{\frac{n_1^2}{L_1^2} + \frac{n_2^2}{L_2^2}}} K_1 \left( 2\pi
n L_3  \sqrt{ \frac{n_1^2}{L_1^2} + \frac {n_2^2} {L_2^2}}
\right).\nn\eeq Higher dimensions could be considered as well by
using the generalization of eq. (\ref{epst}) to higher dimensions
\cite{ambj83-147-1,eliz90-31-170,kirs93-26-2421}.

\section{Zeta function properties on the sphere}
\label{app-sphere}

Let us now turn our attention to the analysis of eq.
(\ref{10}) for the case of the sphere. In $d=2$ this is the same
as the torus and pertinent results can be found above. \\
{\bf d=3:} One easily obtains from eq. (\ref{zetathree}) that
\beq\mbox{Res } \zeta_y^0 \left( \frac 3 2 \right) =0, \quad \quad
\mbox{PP } \zeta_y^0 \left( \frac 3 2 \right) = \pi^2\eeq and from
here eq.
(\ref{10}) can be written down immediately for this example.\\
{\bf d=4:} Here the eigenvalues are $\lambda_l= (l+1)^2$ with
degeneracy ${\rm deg} (l;3) = (l+1) ^2$ and so the relevant zeta
function is
$$\zeta_y ^0 (s) = \zeta_R (2s-2).$$ Therefore everything needed
is available, in particular $$\mbox{Res } \zeta_y^0 \left( \frac 1
2 \right)  = 0, \quad \mbox{PP }\zeta_y^0 \left( \frac 1 2 \right)
= - \frac 1 {12}, \quad \mbox{Res } \zeta_y^0 \left( \frac 3 2
\right) = \frac 1 2 , \quad \mbox{PP }\zeta_y^0 \left( \frac 3 2
\right) = \gamma,$$ and again the final answer for $\zeta ' (0)$
in eq. (\ref{10}) is trivially written down. Also higher
dimensions could be considered along the same lines.

For the more general case of $W(y)=c+(d-2)^2/4$, $c$ an arbitrary
constant, the eigenvalues have the form $$\lambda_l = \left( l
+\frac {d-2} 2 \right)^2 + c.$$ An analytical continuation of the
zeta function $\zeta_y^0 (s)$ in terms of Hurwitz zeta functions
is easily obtained with the help of a binomial expansion in powers
of $c$. This is a well known procedure and we do not present more
details; see, e.g., \cite{eliz94-35-6100,dowk94-35-4989}.


\end{document}